\documentclass[pre,12pt,twocolumn,showpacs]{revtex4}

\usepackage{graphicx}

\begin{document}

\title{Anisotropy in Rupture Lines of Paper Sheets }
\author{I.~L. Menezes-Sobrinho} 
\email{ilima@ufv.br}
 \author{M. S. Couto} 
\author{I. R. B. Ribeiro}
\affiliation{Departamento de F\'{\i}sica - Universidade Federal de Vi\c{c}osa, 36571-000,
Vi\c{c}osa,
Minas Gerais, Brazil}

\begin{abstract}
We have experimentally investigated the fracture process in paper samples submitted to
a uniaxial force. Five types of paper sheets (newsprint, towel, sulfite, silk and couche papers) were fractured along two orthogonal orientations. In
order to characterize the rupture lines of the
paper sheets we utilized the Hurst exponent. Our results indicate a dependence of the Hurst
exponent on the orientation of the paper sheets for samples of
newsprint and, probably, towel and silk papers. For the other types of paper the
Hurst exponent does not depend on the direction of crack propagation.
\end{abstract}


\maketitle

	\section{Introduction}
	\label{intro}
Fracture processes in disordered materials is a subject of intensive research and has attracted
much scientific and industrial interest \cite {livro,zape,maes}. In any rupture experiment the
presence of disorder is fundamental in the dynamics of crack propagation. It is naturally present
in all materials and comes from a variety of different sources. Great theoretical and
experimental efforts have been done trying to understand the process of crack formation and 
propagation in disordered materials. In particular, a significant fraction of
these studies
have been performed from the  point of view of statistical
mechanics, that utilize
concepts such as percolation, fractals, scaling law, etc.
\cite{livro,mand,arc,isma}. 

Over the last decade fibrous materials have been attracting much attention. Several models were
elaborated to describe the fracture process in these
materials \cite{dan,zhang,dux,isma1,isma3,pre,ze,bon,isma2,raul}. However, 
the small number of experimental data does not allow a reliable comparison with the
theoretical results.

 Experiments have shown that the crack line in disordered materials 
can often be described by self-affine scaling \cite{mand,kert,bou}. The surface 
irregularity is often characterized by its roughness, or width, $W(\varepsilon)$, 
defined as the rms value of the fluctuations of the surface height $h_i=h(x_i)$ over a length
scale
$\varepsilon$: 
\begin{equation}
\label{alt}
 W(\varepsilon)=\frac{1}{N}
\sum_{i=\varepsilon}^{N-\varepsilon}\sqrt{\frac{1}{{2\varepsilon+1}}\sum_{j=i-\varepsilon}^{i+\varepsilon}(h(x_j)-\langle
h\rangle_i\rangle)^2}
\end{equation} 
where $\langle
h\rangle_{i}=\frac{1}{{2\varepsilon+1}}\sum_{j=i-\varepsilon}^{i+\varepsilon}h(x_j)$ is the
average height aro\-und the position $i$.

For a self-affine surface, the function $h(x)$ is statistically invariant under an anisotropic
scale transformation. This means that $h(x)$ has the same statistical properties as
$k^{-H}h(kx)$, where $H$ is known as the Hurst (roughness) exponent, which
satisfies $0<H<1$. For
self-affine structures, the
roughness $W$ over a range $\varepsilon$ satisfies the scaling law 
\begin{equation}
W(\varepsilon)\sim \varepsilon^ H.
\label{dim}
\end{equation}
The exponent $H$ does not indicate how rough a surface is. It is a parameter that characterizes
how the roughness, or the variance in the height, depends on the lateral
scale over which it is
measured.

Some experimental works have claimed that the Hurst
exponent $H$ does not depend on the direction of crack propagation and
assumes a universal value of 0.8 \cite{bou,maloy,sch}. However,
this universality was first questioned by Milman {\it et al}. \cite{mil}, who
experimentally found with atomic force microscope measurements of fracture
surface in crystalline
metals a Hurst exponent close to 0.5, a value significantly smaller than 0.8. An exponent
 $H=0.5$ was also observed in materials displaying intergranular fractures, such as
sandstone \cite{boffa}. 
From the theoretical point of view, numerical models have been elaborated in order to evaluate
the Hurst exponent $H$. Computer simulations have  shown that $H\sim 0.7$ in two dimensions 
\cite{hansen,g} and that $H$ ranges from $0.4$ to $0.5$ in
three dimensions \cite{rai,bat}. Nowadays there is a conjecture relating the value of the
Hurst exponent $H$ to the speed of
crack propagation through the sample \cite{bat}. The higher value ($H=0.8)$ has been
associated with a high speed of crack propagation and interpreted as a dynamic
regime. In contrast, the smaller value ($H=0.5)$ was related with a
quasi-static regime, in which the dynamic effects of crack
propagation are
negligible. Experimentally, the smaller and the higher values of $H$ are
connected to the length scale at which the crack is examined \cite
{bou,dag,bou1}. 

Parisi and Caldarelli \cite{parisi} have studied the fracture
surface of three dimensional samples using a model for quasi-static fractures known as Born
Model. They have found a Hurst exponent $H=0.5$, which does not depend on the
direction of crack propagation. 

An experimental procedure used to investigate the fracture process of fibrous materials 
is the rupture of a paper sheet into two parts \cite{kert,ros,sal}. Paper is a good example of a
disordered material due to its structural non-uniformity, which can be confirmed even with the
naked eye. When light is transmitted through it, patches of lighter and darker regions can be
seen. Such darker and lighter parts correspond, respectively, to higher and lower local
densities.
Paper is made from cellulose fibers found in wood, the most frequently used being eucalyptus
and pine. Different types of wood can be utilized to produce cellulose fibers with specific
characteristics, therefore producing papers with different proprieties. 
Generally, the fiber network structure is highly disordered since their fibers are randomly
positioned and oriented. Nevertheless, in some types of paper, the fabrication process is such
that the fibers acquire some order, having a tendency to align along one direction. The fibers
are held together by their ability to form hydrogen bonds with each other. 
The paper strength is a fundamental feature belonging to that group of physical properties 
associated with the paper manufacturing process. Indeed, paper strength is influenced by the
kind, quality, and treatment of
the constituent fibers and by the way the sheet has been formed on the paper machine. 

The analysis of rupture in paper samples is very suitable because the best method to calculate
the Hurst exponent demands the knowledge of the profile, i.e, of the function
$h=f(x)$. In the case of a two-dimensional fracture such a height profile is easily obtained. However, the
evaluation of the Hurst exponent shows a significant dispersion, therefore demanding the analysis of 
a great number of samples. There are in the literature some experiments that investigated the
rupture process in paper samples. 

Maunuksela {\it et al.} \cite{my}, performed experiments on
the dynamics and kinetic roughening of one-dimensional slow-combustion fronts using three types
of paper. They concluded that well-controlled experiments lead to a clear asymptotic scaling
that unequivocally lies in the thermal KPZ universality class.  

Kertész  {\it et al.} \cite{kert} have done experiments using a tensile testing machine to study
the morphology of rupture lines in paper. The samples were submitted to traction and the velocity
of crack propagation was kept constant until the breaking process separated the sample into two
pieces. They have studied five different kinds of paper and found that the Hurst exponent
characterizing the rupture lines did not depend significantly on the type of paper used in the
rupture process. The results obtained indicated a Hurst exponent ranging from $0.63$ to $0.72$,
which did not depend on the direction of crack propagation.

Salminen {\it et al.} \cite{sal1} have studied fracture in paper via acoustic
emission analysis. In
this case, the release of acoustic energy
is related to irreversible deformation, microcraks, and, perhaps, to plasticity.
It has been verified
that the acoustic emission of energy and the waiting times between
acoustic events followed power-law distributions. This remained true while the strain rate was
varied
by more than 2 orders of magnitude. Recently, Salminen {\it et al.} \cite {sal}
have analyzed $6.5~m$
long fracture lines of paper as an example of crack propagation involving
disorder. They have observed that the roughness exhibits a power-law scaling, with an exponent close to $0.60\pm0.1$ for all
samples.

In this paper we have studied the dependence of the rupture line morphologies on the direction of
crack propagation. Several samples of five different types of paper were submitted to a constant
load until their complete rupture. Two directions, perpendicular to each other, were chosen for crack
propagation and the fracture surfaces obtained were characterized by their Hurst exponent. 

The paper is organized as follows. The next section is devoted to present the experimental
procedure used in the rupture process. In section
III, the obtained results are presented and analyzed and, finally, the last section is devoted to
conclusions.

\section{Experimental procedure}
We have carried out rupture experiments using five different types of commercially available papers:
newsprint ($(36.8\pm0.2)~g/m^2$), towel ($(31.7\pm0.1)~g/m^2$), silk ($(51.8\pm0.5)
~g/m^2$), couche ($(82.7\pm0.5)~ g/m^2$) and sulfite ($(78.1\pm0.4)~g/m^2$). Two
directions, $x$ and $y$, perpendicular to each other, were assigned to the paper,
along the width and the length of the sheets respectively. We refer to the $x$ as the
horizontal direction and to the $y$ as the vertical direction (Fig. 1). Square samples
of $20\times20 ~cm$ were cut from the paper sheets and the orientation of the directions $x$ and
$y$ were marked on them. Each sample was fixed at both ends by two parallel clamps, either with
their $x$ or $y$ direction parallel to the clamps. The upper clamp was maintained fixed and in
the lower one a static load $F$ was applied. The same load $F\approx 200~ N$ was used in all
experiments. In order to initiate the propagation of the crack a notch $2~ cm$ long was cut into
the paper in the middle of one of the sides. 

\begin{figure}[hbt]
\begin{center}
\resizebox{8cm}{!}{\includegraphics{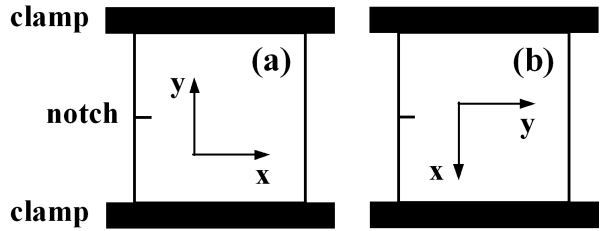}}
\end{center}
\caption{Schematic representation of the experimental apparatus used in the rupture of paper.
 The direction of crack propagation is indicated in each figure. a) horizontal direction
and b) vertical direction.}
\label{sche}
\end{figure}

After fractured, the samples were digitized in a scanner of $450$ dpi
resolution with 256 gray levels and a black background. During the scanning
process the beginning
and the end of the rupture line were discarded. The gray level distribution near
the front of the digitized figure ranges from about 170 on the side belonging to
the paper to about 70 outside of the paper (on a scale that ranges from 0 to
256). The number of colors used to digitize the samples was later decreased to
only 2 (black and white) using the function black and white from
CorelPhoto-Paint 11. The threshold was adjusted until just the smaller fibers
were deleted from the front. The digitized patterns were
analyzed by a home made computer program which allowed us to obtain the
functions $h(x)$
which represent the patterns of the investigated rupture lines. Figure 2 shows
typical functions $h(x)$ obtained for a newsprint sample along the horizontal
and vertical directions.

\begin{figure}[hbt]
\begin{center}
\resizebox{7cm}{!}{\includegraphics{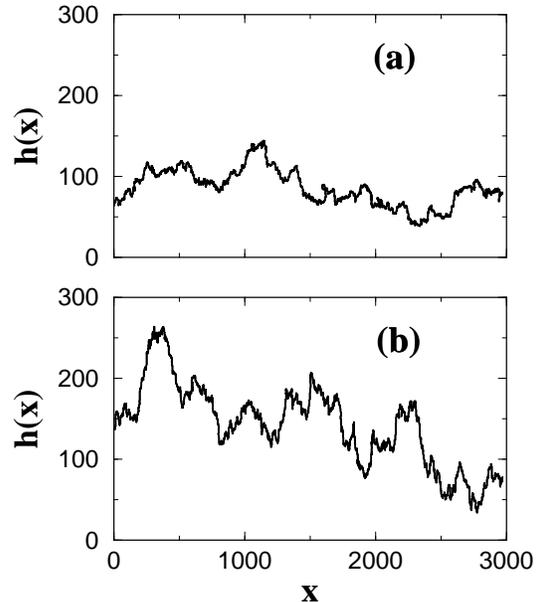}}
\end{center}
\caption{Functions $h(x)$ for two samples of newsprint: (a) horizontal and (b) vertical
direction. The
horizontal and vertical scales are in pixels.}
\label{tear}
\end{figure}

\section{Experimental Results}

From the digitized patterns we evaluated the roughness W (using Eq. (1)) of the rupture lines. A
log-log plot of $W$ versus $\varepsilon$  presents a region where the data have a linear behavior
over at least two decades.
The slope of this region is equal to the Hurst exponent. 

Figure 3 shows typical results for samples of
newsprint, silk and towel papers along the two perpendicular directions. The
slopes of the two straight lines shown are
equal to $0.60$ and $0.72$ for the newsprint, $0.85$ and $0.94$ for the
silk paper and $0.71$ and $0.78$ for the towel paper. The lower value
corresponds to the value of the Hurst exponent along the
horizontal direction and the higher one to the value of the Hurst
exponent along the vertical direction.

\begin{figure}[hbt]
\begin{center}
\resizebox{7cm}{!}{\includegraphics{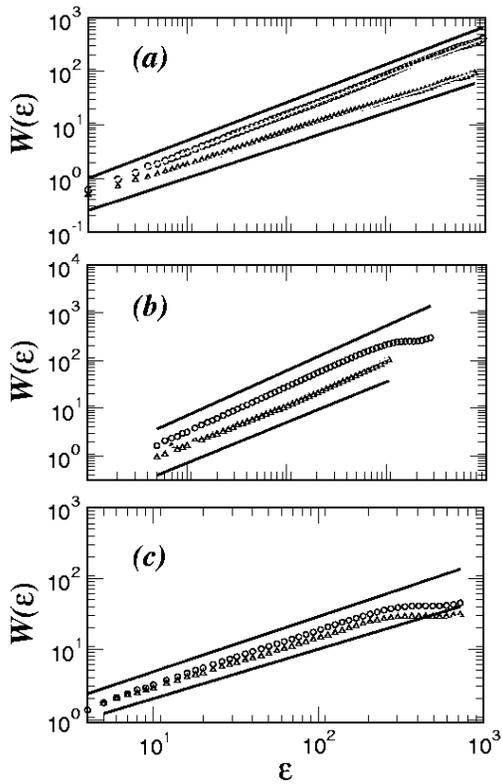}}
\end{center}
\caption{Roughness of rupture lines for two samples of:  (a) newsprint, (b)
silk and (c) towel papers. The rupture lines are oriented along two
different directions: vertical (circles) and horizontal (up triangles). The slopes of the linear
parts allow us to evaluate the Hurst exponents.}
\label{scaling}
\end{figure}

Table I presents the values obtained for the Hurst exponents calculated for each type of paper,
along the horizontal and vertical directions. For all five types of paper, the data
were averaged over 30 statistically independent samples.

\begin{table} 
\begin{center}   
\begin{tabular}{ccc}    \hline\hline
Paper & $H_y$  &$H_x$ \\ \hline
newsprint & $0.73\pm 0.02$ &$ 0.65\pm 0.02$ \\
couche paper & $0.74\pm0.04$ & $0.74\pm0.03$ \\
towel paper & $0.78\pm 0.02$ & $0.71\pm 0.03$ \\
sulfite paper & $0.80\pm0.04$ & $0.77\pm 0.04$ \\ 
silk paper & $0.94\pm0.02$ & $0.90\pm0.01$ \\
\hline\hline
\end{tabular}
\end{center}
\caption{Hurst exponent for different kinds of paper sheet obtained along two
directions ($H_y=$ Hurst exponent for the vertical direction and $H_x=$ Hurst exponent for the
horizontal direction).}
\end{table}

Our results indicate a dependence of the Hurst exponent upon the direction of
crack propagation
for the samples of newsprint, towel and, probably, silk papers. This dependence is
related to the inhomogeneity
of the paper. For these three types of paper the Hurst exponent is larger along the vertical
direction than along the horizontal one. This indicates that in these papers, the propagation speed  of
a crack should be larger along the vertical direction. 
  
For the other two types of paper it was observed that, within the experimental error, the Hurst
exponent of the studied samples did not depend on the direction of crack propagation. Notice
that in Table I the values obtained for the Hurst exponents along both
directions are higher
than $0.5$. This result indicates that the fracture process investigated here can be identified
as belonging to the dynamic regime and, also, corresponds to measurements at large length
scales \cite{dag}.  

The results in Table I, for the newsprint, clearly reveal that the Hurst exponent depends on the direction along which it is measured. In order to confirm this
result we have increased the scanner resolution to 800 dpi and used three other test methods for the crack roughness
analysis: max-min difference \cite {max}, power spectrum \cite{power} and best linear least-square fitting \cite{jaf}.

The max-min method consists of the computation of $h_{max}(\epsilon)$, defined as the difference between the lowest and the highest heights $h$ within
a certain window of size $\epsilon$, averaged over all possible origins $x$
of the window. For a self-affine profile, $h_{max}$ scales as a
function of $\epsilon$  according to 
\begin{equation}
h_{max}\sim\epsilon^H. 
\end {equation}

The power spectrum method measures the power spectrum of the interface, and not
the height $h(x)$. In this case it is calculated the structure factor, or power
spectrum 
\begin{equation}
(\vec{k})=\langle h(\vec{k})h(-\vec{k})\rangle~,
\end{equation}
 where
\begin{equation}
h(\vec{k})=\frac{1}{L^{1/2}}\sum_{\vec{x}}[h(\vec{x})-\langle
h\rangle_i\rangle]exp(i\vec{k}.\vec{x})~,
\end{equation}
and $k$ is the wave factor.
The structure factor scales, for a self-affine profile, as 
\begin{equation}
s(k)\sim k^{-(2H+1)}. 
\end{equation}

In the method of the best linear least-square fitting, the roughness $W(\epsilon)$ over a lenght
scale $\epsilon $ is given by
\begin{equation}
 W(\epsilon)={1\over N}\sum_{i=1}^{N}w_i(\epsilon)
\label{rug}
\end{equation}
and the local roughness $w_i(\epsilon)$ is defined as
\begin{equation}
w_i^2(\epsilon)={1\over(2\epsilon +1)}\sum_{j=i-\epsilon}^{i+\epsilon}
[h_j-(a_i(\epsilon)x_j+b_i(\epsilon))]^2.
\end{equation}
$a_i(\epsilon)$ and $b_i(\epsilon)$ are the linear fitting coefficients to the
displacement ratio data on the interval $[i-\epsilon,i+\epsilon]$ centered at
the position $i$. For a self-affine profile, the roughness $W$ of the fracture surface satisfies the scaling law
\begin{equation}
W(\epsilon)\sim\epsilon^H.
\end{equation}

The values of the Hurst exponents, estimated using the four methods, for
the newsprint are given in Table II. Notice that the three methods agree on the
values of the Hurst exponents. This fact indicates that the crack surfaces are
truly
self-affine and confirms the conjecture that the Hurst exponent depends on the
direction of fracture. 

\begin{table} 
\begin{center}   
\begin{tabular}{ccc}    \hline\hline
method & $H_y$  &$H_x$ \\ \hline
rms & $0.69\pm 0.02$ &$ 0.61\pm 0.02$ \\
max-min & $0.73\pm0.02$ & $0.63\pm0.02$ \\
power spectrum & $0.71\pm 0.03$ & $0.63\pm 0.03$ \\
best linear least-square fitting & $0.68\pm 0.03$ & $0.58\pm 0.03$ \\
\hline\hline
\end{tabular}
\end{center}
\caption{Hurst exponent obtained for newsprint using different methods.}
\end{table}

Figures 4(a) and 4(b) show optical microphotographs of newsprint and couche paper samples, respectively.
For the newsprint it can be clearly seen that there are many long fibers running, preferentially,
along the horizontal direction of the figure. This orientation coincides with the horizontal
direction of the paper. For the couche paper the fibers are also long but they do not appear to
be oriented along any preferential direction, being more intertwined than those of the newsprint.
The orientation of the fibers is produced during the paper fabrication process and seems to play
an important role in the roughness of the rupture lines. The actual mechanism of fracture is not
yet completely understood at the microscopic level. For the paper to be
fractured, the rupture of the
hydrogen bonds between the fibers must occur in order that they can be separated from one
another. However, the rupture of individual fibers seems to be not necessary. 
When the newsprint is fractured along the horizontal direction the crack can propagate mainly
through the separation of one fiber from another by breaking the hydrogen bonds. The rupture of
individual fibers, if it occurs at all, does not have to happen very often since most of the
fibers are aligned parallel to the direction of crack propagation. In this case, we can suppose
that the rupture process can occur due to two different possible fracture modes: tensile
fracture of the hydrogen bonds between the fibers and shear fracture of the fibers. When the
newsprint is fractured along the vertical direction the
crack has to propagate in a direction perpendicular to the fibers. In this case, the hydrogen
bonds between the fibers are mainly subjected to shear fracture mode while the fibers are mainly
subjected to tensile fracture mode. Again the rupture of individual fibers is not necessary. If
the hydrogen bonds between them break, they can simply slide past each other. The separation of
the fibers is probably more difficult in this process as compared to that for the horizontal
direction since, as the fibers slide past each other, hydrogen bonds that were broken can be
formed again. Rupture of individual fibers should be even more difficult since the force
necessary to rupture one fiber must be greater than that to break a hydrogen bond. Consequently,
the crack has to change its direction of propagation many times, trying to find a weaker point to
fracture and the rupture line formed is rougher than that formed along the horizontal direction.
Microscopic examination of the rupture lines show that there are much more long
fibers protruding
from the paper when the paper is fractured along the vertical direction than when it is fractured
along the horizontal direction. This indicates that, indeed, the fibers are just being pulled
apart without being fractured. Unfortunately, comparison of the end of the protruding fibers
with the ends of the fibers away from the rupture line could not reveal if the former where
broken during the fracture process. It is possible to make an analogy between paper fracture and
microscopic failure mechanisms of a unidirectional composite material \cite {livro} if the
hydrogen bonds are considered as being the matrix.

For the other types of paper that presented a dependence of the Hurst exponent on the orientation
of the paper sheet, it was also possible to see the same kind of preferential fiber orientation, although much less pronounced than for the newsprint. This, probably, is the reason why
the difference between $H_x$ and $H_y$ for the towel and the silk papers is smaller than for
the newsprint. For the two types of paper that did not show dependence between the Hurst exponent
and
the orientation of the paper sheet it was not possible to detect any orientation of the fibers.
There seems to be, also, a relation between the size of the fibers and the value of the Hurst
exponent. Papers composed of long fibers have smaller exponents than those of papers composed
of short fibers. 
\begin{figure}[hbt]


\caption{Optical microphotographs for: (a) newsprint and (b) couche paper. Both papers are
oriented with their $y$ direction along the length of the page.}
\label{lines}
\end{figure}

\section{Conclusion}
\label{discussion}
In this paper we have experimentally investigated the fracture process in five different types of
paper submitted to a static load. The rupture line of the samples were characterized by the Hurst
exponent, calculated along two orthogonal orientations of the paper: vertical and
horizontal. This procedure was utilized to investigate the effect of anisotropy on the Hurst
exponent.

For the samples of newsprint our results clearly indicate that the
Hurst exponent depends on the orientation of the paper. This dependence is associated with the
orientation of the fibers within the paper, which is determined by its
fabrication process. For the samples of silk and towel papers the difference
between the Hurst exponents is small, almost within the experimental error,
indicating a probable 
dependence of the Hurst exponent upon the orientation of the paper. 

For all paper samples studied the rupture process is very rapid. This can be verified by the
value of the Hurst exponent, higher than $0.5$ for all samples tested. Therefore, the
fracture process can be classified as a dynamic one. 

Contrary to the results reported on the literature, our results support the idea that the value
of the Hurst exponent for paper sheets is not
universal. It depends on both the type of paper used and, more interestingly, it can also depend
on the direction along which it is measured for the same paper sheet.

\medskip
\centerline{\bf Acknowledgments}

We thank  M.L Martins and J. G. Moreira for helpful criticism of the manuscript. We also thank J.
A. Redinz and S. C. Ferreira Jr. for many valuable discussions. This work was partially supported by CNPq
(Brazilian agency).   
 
%

\end{document}